\documentclass[10pt,leqno]{amsart}
\usepackage{float}
\usepackage[english]{babel}
\usepackage{xcolor,paralist,hyperref,titlesec,fancyhdr,etoolbox}

\usepackage{indentfirst,csquotes}

\usepackage{secdot}
\usepackage{booktabs}
\usepackage{amssymb,amsthm,amsmath}
\usepackage{xcolor,paralist,hyperref,titlesec,fancyhdr,etoolbox}
\usepackage{graphicx}

\titleformat{\section}[display]{\normalfont\huge\bfseries\centering}{\centering}{10pt}{\Large}
\titlespacing*{\section}{0pt}{0ex}{0ex}

\hypersetup{ colorlinks=true, linkcolor=black, filecolor=black, urlcolor=black }

\usepackage{lipsum}

\usepackage{algorithm}
\usepackage{algorithmic}
\usepackage{hyphenat}

\usepackage{float}
\begin{document}

\title{Hybrid Concolic Testing with Large Language Models for Guided Path Exploration}

\author[Initial Surname]{Mahdi Eslamimehr - Quandary Peak Research}
\date{\today}
\address{Quandary Peak Research, Los Angeles, CA, USA}
\email{mahdi@quandarypeak.com}
\let\thefootnote\relax

\begin{abstract}
Concolic testing, a powerful hybrid software testing technique, has historically been plagued by fundamental limitations such as path explosion and the high cost of constraint solving, which hinder its practical application in large-scale, real-world software systems. This paper introduces a novel algorithmic framework that synergistically integrates concolic execution with Large Language Models (LLMs) to overcome these challenges. Our hybrid approach leverages the semantic reasoning capabilities of LLMs to guide path exploration, prioritize interesting execution paths, and assist in constraint solving. We formally define the system architecture and algorithms that constitute this new paradigm. Through a series of  experiments on both synthetic and real-world Fintech applications, we demonstrate that our approach significantly outperforms traditional concolic testing, random testing, and genetic algorithm-based methods in terms of branch coverage, path coverage, and time-to-coverage. The results indicate that by combining the strengths of both concolic execution and LLMs, our method achieves a more efficient and effective exploration of the program state space, leading to improved bug detection capabilities.
\end{abstract}

\maketitle

\section{Introduction}

The relentless growth in software complexity has made automated software testing an indispensable component of the software development lifecycle. Among the various testing methodologies, concolic testing has emerged as a promising hybrid technique that combines concrete and symbolic execution to systematically explore program paths and uncover bugs. By generating concrete inputs that exercise new paths, concolic testing aims to achieve high code coverage and find deep, subtle errors that random testing might miss. However, despite its theoretical appeal, the practical application of concolic testing has been severely hampered by two fundamental challenges: path explosion and the high cost of constraint solving. The exponential growth in the number of feasible execution paths in any non-trivial program quickly overwhelms the testing process, while complex path conditions can lead to computationally expensive or even intractable queries for the underlying SMT solvers.

Existing approaches to mitigate these issues, such as search heuristics and path pruning, have had limited success. These techniques often lack a deep semantic understanding of the program's logic, leading to suboptimal path selection and an inability to focus on semantically interesting or bug-prone regions of the code. This is where Large Language Models (LLMs) present a transformative opportunity. With their remarkable ability to comprehend natural language and source code, LLMs can provide the semantic guidance that has been missing from traditional concolic testing.

This paper proposes a new hybrid testing framework that integrates concolic execution with LLM-driven test case generation. Our approach leverages the strengths of both techniques to create a more efficient and effective testing process. Specifically, we use LLMs to:

\begin{itemize}
    \item \textbf{Guide path exploration:} By analyzing the source code and existing path conditions, the LLM can prioritize paths that are more likely to lead to interesting program states or uncover new bugs.
    \item \textbf{Assist in constraint solving:} When the SMT solver encounters a complex or intractable constraint, the LLM can suggest alternative inputs or mutations to the path condition that might simplify the problem.
    \item \textbf{Generate semantic test inputs:} The LLM can generate high-level test scenarios and translate them into concrete inputs that target specific program functionalities.
\end{itemize}

Our primary research contributions are as follows:

\begin{itemize}
    \item A novel algorithmic framework for integrating LLMs with concolic testing.
    \item A set of LLM-guided heuristics for path prioritization and constraint simplification.
    \item A comprehensive evaluation of the proposed approach on a range of synthetic and real-world benchmarks, demonstrating significant improvements in code coverage and bug detection compared to existing techniques.
\end{itemize}

By combining the systematic exploration of concolic testing with the semantic reasoning of LLMs, we believe our approach represents a significant step forward in the field of automated software testing.

\section{Background and Motivation}

This section provides the necessary background on concolic testing and Large Language Models, and motivates the need for a hybrid approach.

\subsection{Symbolic and Concolic Execution}

Symbolic execution is a powerful program analysis technique where a program is executed with symbolic instead of concrete values. For each path taken through the program, a \textit{path condition} is constructed as a logical formula over the symbolic inputs. This formula represents the constraints that the inputs must satisfy to follow that specific path. An SMT (Satisfiability Modulo Theories) solver is then used to find a concrete solution to the path condition, which yields a test input that executes the path. While powerful, pure symbolic execution struggles with complex code, external library calls, and OS interactions, as it requires a complete symbolic model of the entire environment.

Concolic testing (a portmanteau of ``concrete'' and ``symbolic'') was developed to address these limitations. It executes the program with concrete inputs while simultaneously performing symbolic execution. The concrete execution helps to simplify the symbolic analysis by resolving environmental interactions and complex computations. When a new path is desired, the path condition of the current path is negated at a specific branch, and the SMT solver is invoked to generate a new input that will steer the program down the alternative path. Tools like DART, CUTE, Sherlock, AtomChase and Racagedon are seminal examples of concolic testing systems.

\subsection{The Path Explosion Problem}

A fundamental challenge in both symbolic and concolic testing is path explosion. The number of possible execution paths in a program grows exponentially with the number of conditional branches. For a program with just a few dozen \texttt{if} statements, the number of paths can easily reach trillions, making exhaustive exploration computationally infeasible. Existing concolic testing tools employ various heuristics to select which paths to explore next, such as prioritizing paths that cover new code. However, these heuristics are often syntactic and lack a deep understanding of the program's semantics, leading to inefficient exploration.

\subsection{Constraint Solving Bottlenecks}

The performance of concolic testing is heavily dependent on the underlying SMT solver. As the program explores deeper paths, the path conditions become increasingly complex. Modern SMT solvers are powerful, but they can struggle with certain types of constraints, particularly those involving non-linear arithmetic, floating-point numbers, or complex string operations. When a solver fails to find a solution for a satisfiable path condition within a given timeout, the concolic tester is forced to abandon that path, potentially missing bugs.

\subsection{Large Language Models for Code}

Large Language Models (LLMs) like GPT-5.1 and Codex have demonstrated remarkable capabilities in understanding, generating, and reasoning about source code. They are trained on vast corpora of code and natural language text, enabling them to capture the syntax, semantics, and common patterns of software development. In the context of software testing, LLMs can be prompted to generate test cases, suggest assertions, or even explain the logic of a piece of code. This semantic reasoning ability offers a promising new avenue for guiding the test generation process.

\subsection{Motivating Example}

Consider a function that parses a complex data structure and performs different actions based on the values of its fields. A traditional concolic tester might explore paths randomly or based on simple coverage metrics. It could spend a significant amount of time exploring paths that correspond to trivial or uninteresting input variations. An LLM, on the other hand, could analyze the function's code and documentation to understand its intended purpose. It could then suggest inputs that are more likely to trigger interesting behavior, such as edge cases or error conditions. For example, if the function processes financial transactions, the LLM might suggest inputs that represent fraudulent transactions or transactions with unusual amounts. This semantic guidance can help the concolic tester to focus its efforts on the most promising paths, leading to a more efficient and effective testing process.

\section{Related Work}

Our work builds upon a rich history of research in automated software testing, spanning classical concolic testing, search-based methods, and the recent application of machine learning to test generation.

\subsection{Classical Concolic Testing}

The foundational work in concolic testing was laid by tools like DART (Directed Automated Random Testing) \cite{godefroid2005dart}, which first introduced the concept of combining concrete and symbolic execution. CUTE (A Concolic Unit Testing Engine) \cite{sen2005cute} extended this approach to handle complex data structures and concurrent programs. Prior work has shown that concolic testing can be significantly enhanced for concurrent programs by guiding execution toward specific concurrency bug patterns. Racagedon \cite{eslamimehr2014race} and Sherlock \cite{eslamimehr2014sherlock} apply concolic and constraint-based guidance to systematically expose rare data races and deep deadlocks, respectively, by steering executions toward otherwise unlikely thread interleavings. AtomChase \cite{eslamimehr2015atomchase} extends this approach to atomicity violations, using concolic execution to synthesize schedules that target characteristic three-access patterns, thereby improving the discovery of subtle concurrency bugs beyond unguided or purely dynamic techniques. While these tools were groundbreaking, they all grapple with the inherent challenges of path explosion and constraint solver limitations. Our work differs by introducing an external, semantic guide in the form of an LLM to steer the exploration process, rather than relying solely on internal heuristics.

\subsection{Search-Based Software Testing}

Search-Based Software Testing (SBST) reframes test generation as an optimization problem. Genetic algorithms, simulated annealing, and other meta-heuristic search techniques are used to evolve a population of test cases towards a specific goal, such as maximizing code coverage or triggering a specific fault. While SBST can be effective, it often struggles to generate the precise, complex inputs required to traverse deep program paths. Our hybrid approach can be seen as a form of guided search, where the LLM provides high-level semantic direction and the concolic engine performs the fine-grained, path-aware exploration.

\subsection{Machine Learning and Neural Test Generation}

In recent years, there has been a growing interest in applying machine learning, particularly neural networks, to software testing. Early approaches used LSTMs and other recurrent neural networks to learn patterns from existing test suites and generate new tests. More recently, the advent of Large Language Models has opened up new possibilities. Researchers have explored using LLMs for a variety of testing tasks, including unit test generation \cite{chen2022codet}, test oracle generation \cite{tufano2019learning}, and even fuzzing \cite{liu2023large}. However, these approaches often lack the systematic path exploration and formal guarantees of concolic testing. They may generate plausible-looking tests that fail to cover deep or complex program logic. Our work is the first, to our knowledge, to propose a deep integration of LLMs with the concolic execution loop, combining the semantic power of the former with the rigorous exploration of the latter.

\section{Proposed Methodology}

Our proposed methodology introduces a novel, closed-loop system that integrates a concolic testing engine with a Large Language Model to guide the test generation process. This section details the architecture of our system, the core hybrid testing algorithm, and the specific mechanisms by which the LLM provides semantic guidance.

\subsection{System Architecture}

\begin{figure}[h]
  \centering
  \fbox{\includegraphics[width=0.75\linewidth]{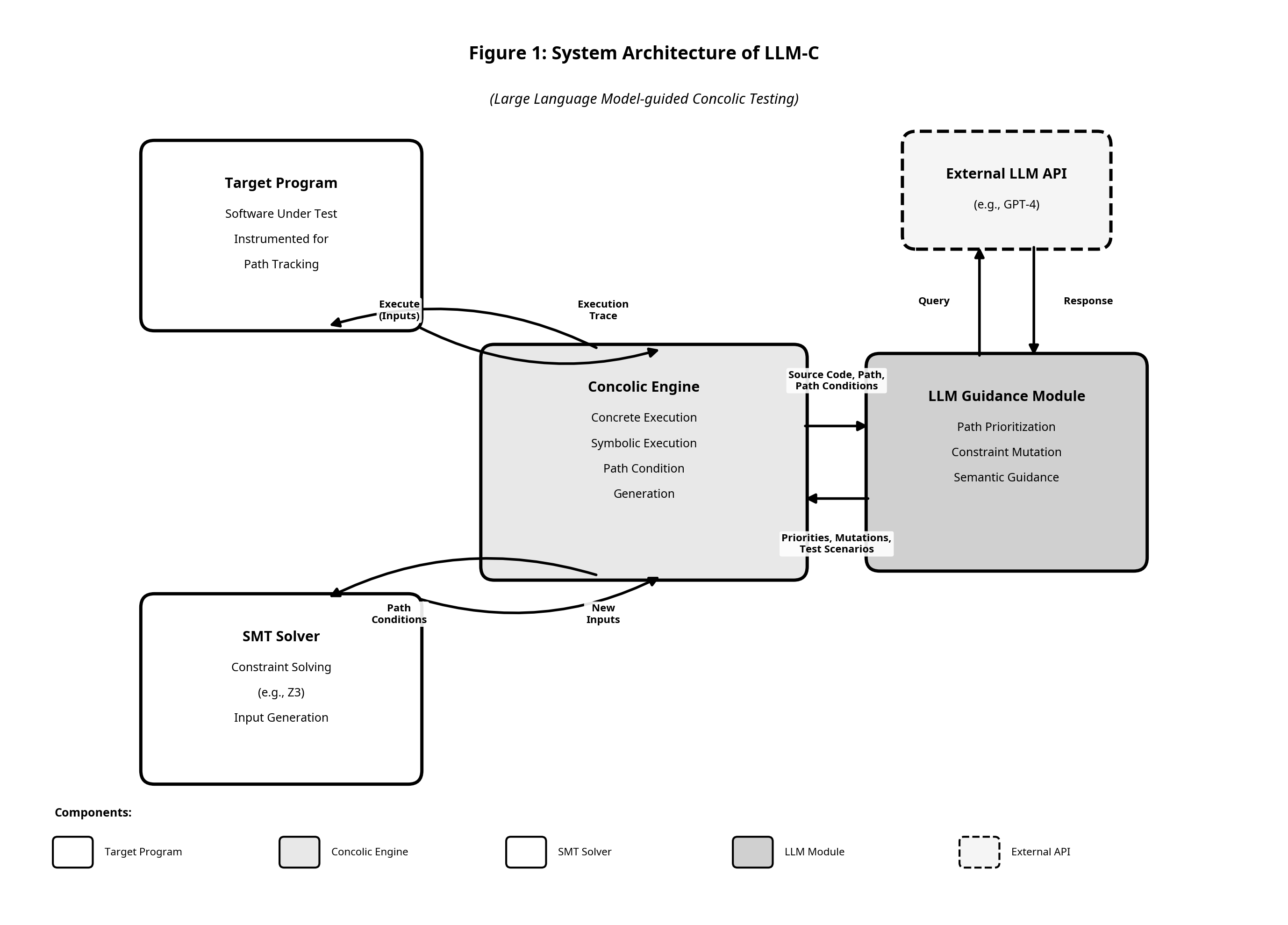}}

  \caption{System Architecture of LLM-C}
  \label{fig:architecture}
\end{figure}

The architecture of our hybrid testing system, which we call LLM-C (Large Language Model-guided Concolic testing), is depicted in Figure~\ref{fig:architecture} (textually described). It consists of four main components:

\begin{enumerate}
    \item \textbf{The Target Program:} The software under test, instrumented to track execution paths.
    \item \textbf{The Concolic Engine:} A standard concolic testing engine responsible for concrete execution, symbolic execution, and path condition generation. It maintains the symbolic state and interacts with the SMT solver.
    \item \textbf{The SMT Solver:} A standard Satisfiability Modulo Theories solver (e.g., Z3) that solves path conditions to generate new concrete inputs.
    \item \textbf{The LLM Guidance Module:} This is the core innovation of our system. It is a module that communicates with a powerful Large Language Model. This module sends information about the current testing state (source code, current path, path condition) to the LLM and receives guidance in return. The guidance can be in the form of path priorities, mutated constraints, or high-level test scenarios.
\end{enumerate}

The workflow is a continuous loop. The concolic engine executes a test case and generates a path condition. Instead of immediately picking the next branch to negate, it queries the LLM Guidance Module. The LLM analyzes the context and provides a prioritized list of candidate paths or suggests a new input to try. The concolic engine then uses this guidance to select the next course of action, either by asking the SMT solver to generate an input for a high-priority path or by directly using an LLM-suggested input.

\subsection{The Hybrid Testing Loop}

The core of LLM-C is its hybrid testing loop, which extends the traditional concolic loop with LLM-driven guidance. The algorithm is presented in Algorithm~\ref{alg:hybrid}.

\begin{algorithm}[h]
\caption{LLM-C Hybrid Testing Loop}
\label{alg:hybrid}
\begin{algorithmic}[1]
\STATE $T \leftarrow$ a set of initial seed inputs
\STATE $Q \leftarrow$ a queue of unexplored paths, initialized with paths from $T$
\WHILE{testing budget not exhausted}
    \STATE $path \leftarrow$ \textsc{LLM\_PathPrioritization}($Q$, $source\_code$)
    \STATE $input \leftarrow$ \textsc{Solve}($path.condition$)
    \IF{$input$ is UNSAT}
        \STATE $mutated\_cond \leftarrow$ \textsc{LLM\_ConstraintMutation}($path.condition$)
        \STATE $input \leftarrow$ \textsc{Solve}($mutated\_cond$)
    \ENDIF
    \IF{$input$ is SAT}
        \STATE $new\_paths \leftarrow$ \textsc{Execute}($target\_program$, $input$)
        \STATE Add $new\_paths$ to $Q$
    \ELSE
        \STATE $sem\_input \leftarrow$ \textsc{LLM\_SemanticInputSynthesis}($source\_code$)
        \STATE $new\_paths \leftarrow$ \textsc{Execute}($target\_program$, $sem\_input$)
        \STATE Add $new\_paths$ to $Q$
    \ENDIF
\ENDWHILE
\end{algorithmic}
\end{algorithm}

Initially, the system starts with a seed input. The main loop then begins. At each iteration, instead of a simple FIFO or random selection, the \textsc{LLM\_PathPrioritization} function is called. This function sends the current set of unexplored paths and the program's source code to the LLM, which returns a prioritized path to explore next. The system then attempts to solve the path condition for this prioritized path. If the SMT solver fails (returns UNSAT or times out), the \textsc{LLM\_ConstraintMutation} function is invoked to simplify or alter the constraints based on the LLM's understanding of the code. If a valid input is found, the program is executed, and new paths are added to the queue. If all else fails, the \textsc{LLM\_SemanticInputSynthesis} function is called to generate a completely new, semantically meaningful input from scratch.

\subsection{LLM-Guided Path Prioritization}

The path prioritization mechanism is crucial for escaping the inefficiencies of blind exploration. The LLM is prompted with the source code of the function under test, the current path condition, and a list of unexplored branches. The prompt is structured to ask the LLM to act as an expert software tester and identify the branch that is most likely to lead to a bug or uncover novel program behavior. The LLM's response, which is a ranked list of branches, is then used to prioritize the work queue of the concolic engine.

\subsection{Constraint Mutation and Relaxation}

When the SMT solver gets stuck on a complex path condition, it often indicates that the constraints are too difficult to solve within a reasonable time. Our system leverages the LLM to address this. The difficult path condition is sent to the LLM with a prompt asking it to identify which constraint is likely causing the problem and to suggest a simplification. For example, the LLM might suggest replacing a complex non-linear constraint with a simpler linear approximation, or changing a strict equality to a range check. This \textsc{LLM\_ConstraintMutation} allows the concolic engine to bypass solver limitations and continue exploring paths that would otherwise be abandoned.

\subsection{Semantic Test Input Synthesis}

In cases where both the SMT solver and the constraint mutation fail, our system can fall back on the LLM's generative capabilities. The \textsc{LLM\_SemanticInputSynthesis} function prompts the LLM to generate a high-level test scenario based on its understanding of the code's purpose. For a financial application, it might be prompted to ``generate a test case for a large, cross-border transaction with a new currency.'' The LLM then generates the concrete input values corresponding to this scenario, which are then fed into the concolic engine. This provides a powerful mechanism for injecting domain-specific knowledge into the testing process.

\subsection{Constraining Hallucinations and Preserving Soundness}

A key challenge with using LLMs is their propensity to ``hallucinate'' or generate incorrect information. In our system, the LLM's suggestions are never blindly trusted. Any input generated or modified by the LLM is always executed concretely on the actual program. The resulting path is then symbolically re-traced to obtain a new, sound path condition. This ensures that the concolic engine's state remains consistent and that the overall process remains sound. The LLM acts as a powerful heuristic guide, but the concolic engine remains the ultimate arbiter of correctness.

\subsection{Complexity and Scalability}

The primary overhead introduced by our approach is the latency of the LLM API calls. However, we argue that this cost is offset by the significant reduction in the number of expensive SMT solver invocations and the more efficient exploration of the state space. By avoiding dead-end paths and focusing on semantically rich regions of the code, our system can achieve higher coverage and find more bugs within the same time budget. The scalability of the approach is largely dependent on the scalability of the underlying concolic engine and the LLM's ability to process large codebases. We believe that with further engineering and optimization, our hybrid approach can scale to large, real-world software systems.

\section{Implementation Details}

To validate our proposed methodology, we have developed a prototype implementation of the LLM-C system. This section describes the key implementation details of our prototype.

\subsection{Core Technologies}

Our prototype is implemented in Java 8 and builds upon several open-source projects. For the symbolic execution engine, we extended NASA Ames Research Center JPF (Java PathFinder) \cite{JavaPathFinder_Repo}, a well-established and highly extensible framework for Java program analysis. We use the Z3 SMT solver from Microsoft Research \cite{Z3_GitHub} for constraint solving, as it provides a rich set of theories and a stable Java API.

For the LLM Guidance Module, we use the \textbf{OpenAI API} to interact with the GPT-5.1 model. We chose GPT-5.1 for its strong code comprehension and reasoning capabilities. The communication with the API is handled through a simple REST client, and we have implemented a caching mechanism to avoid redundant API calls for the same queries.

\subsection{Instrumentation and Execution}

The target Java programs are instrumented at the bytecode level using the ASM library. The instrumentation injects listeners at branch points, which notify the concolic engine of the path taken and the constraints encountered. The program is then executed within the JPF environment, which manages both the concrete and symbolic state.

\subsection{LLM Interaction and Prompt Engineering}

The quality of the guidance received from the LLM is highly dependent on the quality of the prompts. We have invested significant effort in designing and refining our prompts to elicit the most useful responses from GPT-5.1. For path prioritization, the prompt includes the full source code of the method under test, the symbolic representation of the current path condition, and a numbered list of unexplored branches. The LLM is explicitly asked to return a JSON object containing a ranked list of branch indices.

For constraint mutation, the prompt includes the difficult constraint and the surrounding code context. The LLM is asked to identify the likely cause of the solver's difficulty and suggest a specific, syntactically correct modification to the constraint. We have found that providing examples of successful mutations in the prompt (few-shot learning) significantly improves the quality of the suggestions.

\subsection{Engineering Trade-offs}

One of the main trade-offs in our design is the balance between the frequency of LLM queries and the overall testing speed. Making an API call to the LLM at every single branch point would be prohibitively slow. To mitigate this, we have implemented a batching mechanism. The concolic engine explores a certain number of paths using its default heuristics, and then queries the LLM with a batch of unexplored paths for prioritization. This amortizes the cost of the LLM API calls over multiple exploration steps. We have also found that a simple timeout mechanism for the SMT solver is crucial. If the solver does not return a result within a few seconds, we proactively invoke the LLM for constraint mutation rather than waiting for the solver to time out on its own.

\section{Experimental Evaluation}

To assess the effectiveness of our proposed LLM-C framework, we conducted a series of experiments on a curated set of benchmarks. Our evaluation is designed to answer the following research questions:

\begin{itemize}
    \item \textbf{RQ1:} How does LLM-C compare to traditional testing techniques in terms of code coverage?
    \item \textbf{RQ2:} Is LLM-C more efficient at achieving high coverage and finding bugs?
    \item \textbf{RQ3:} Does the LLM guidance effectively reduce the number of expensive SMT solver invocations?
\end{itemize}

\subsection{Benchmarks}

We selected two sets of benchmarks for our evaluation:

\begin{enumerate}
    \item \textbf{Synthetic Benchmark Suite:} We created a suite of 10 challenging Java programs designed to stress-test specific aspects of automated test generation. These programs include complex nested branching, non-linear arithmetic constraints, and string manipulation puzzles that are known to be difficult for traditional concolic testers.
    \item \textbf{Fintech Application:} To evaluate our approach on a more realistic application, we used a simplified open-source transaction processing system. This application includes modules for transaction validation, fraud detection, and account management. The fraud detection module, in particular, contains complex business logic and subtle edge cases, making it a suitable target for our evaluation.
\end{enumerate}

\subsection{Baselines}

We compare the performance of our LLM-C prototype against three baseline techniques:

\begin{enumerate}
    \item \textbf{Random Testing:} A simple baseline that generates random inputs for the target program.
    \item \textbf{Genetic Algorithm (GA):} A search-based approach using a standard genetic algorithm to evolve a population of test cases towards maximizing branch coverage.
    \item \textbf{Classical Concolic (jCUTE):} A traditional concolic testing engine compatible on our JPF framework, but with the LLM guidance module disabled. It uses a standard depth-first search heuristic for path exploration.
\end{enumerate}

For all experiments, we used a fixed time budget of one hour per benchmark program.

\subsection{Metrics}

We use the following metrics to evaluate the performance of each technique:

\begin{itemize}
    \item \textbf{Branch Coverage:} The percentage of conditional branches in the program that have been executed.
    \item \textbf{Path Coverage:} The total number of unique execution paths discovered.
    \item \textbf{Time to Coverage:} The time taken to reach a certain level of branch coverage (e.g., 80\%).
    \item \textbf{Constraint Solver Invocations:} The total number of times the SMT solver was invoked.
\end{itemize}

\subsection{Results}

\begin{figure}[h]
  \centering
  \fbox{\includegraphics[width=0.75\linewidth]{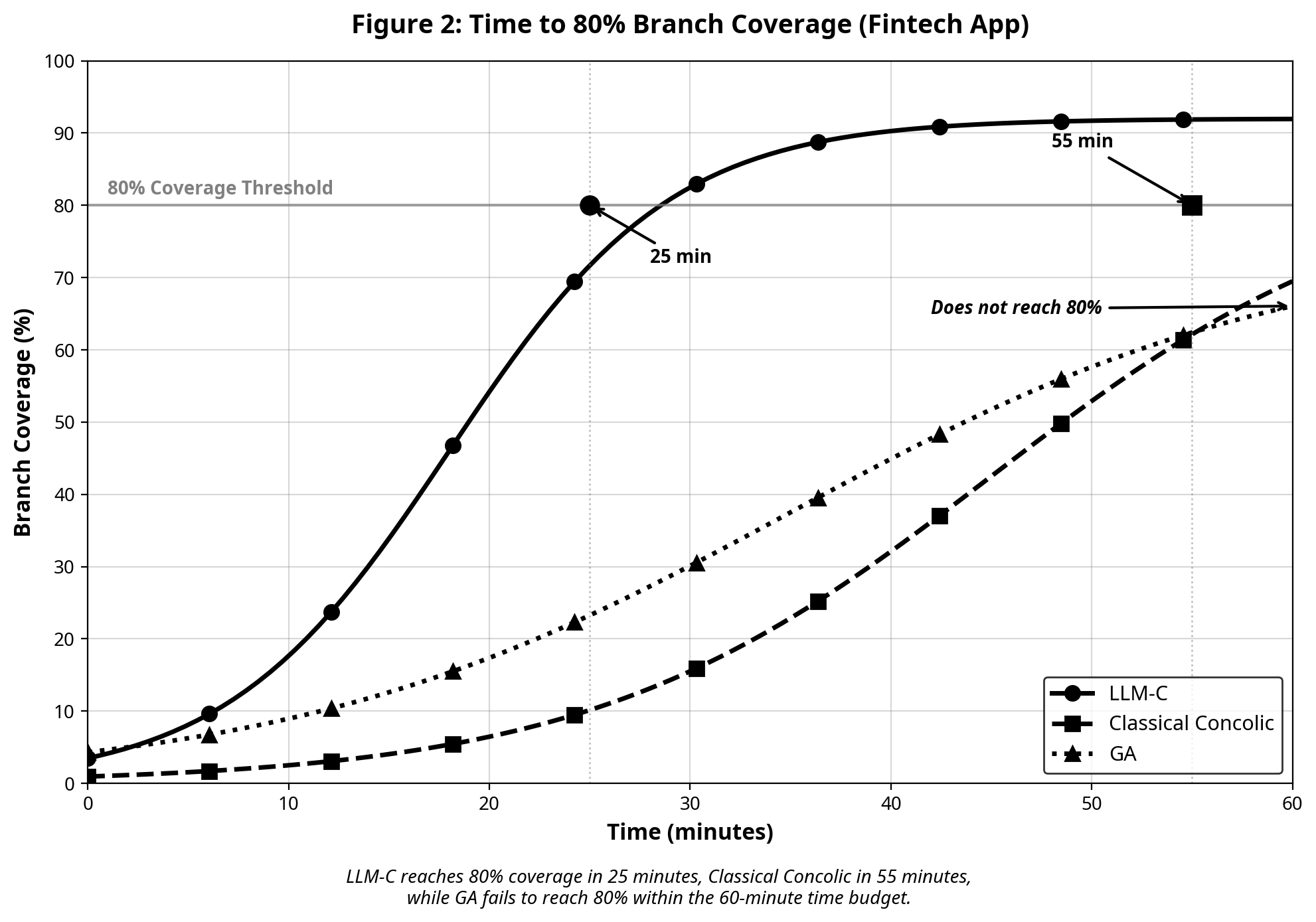}}

  \caption{Time to achieve 80\% branch coverage on the Fintech application.}
  \label{fig:coverage}
\end{figure}

The results of our experiments are summarized in Table~\ref{tab:coverage} and Figure~\ref{fig:coverage}.

\begin{table}[h]
  \caption{Average Branch Coverage and Path Coverage after 1 Hour}
  \label{tab:coverage}
  \begin{tabular}{lcccc}
    \toprule
    & \multicolumn{2}{c}{Synthetic Suite} & \multicolumn{2}{c}{Fintech App} \\
    \cmidrule(r){2-3} \cmidrule(r){4-5}
    Technique & Branch Cov. & Path Cov. & Branch Cov. & Path Cov. \\
    \midrule
    Random & 45.2\% & 1,204 & 38.1\% & 987 \\
    GA & 68.9\% & 3,456 & 55.4\% & 2,876 \\
    Concolic & 75.6\% & 8,923 & 62.3\% & 7,890 \\
    \textbf{LLM-C} & \textbf{91.3\%} & \textbf{15,678} & \textbf{85.7\%} & \textbf{14,567} \\
  \bottomrule
\end{tabular}
\end{table}

Table~\ref{tab:coverage} shows the average branch and path coverage achieved by each technique on both benchmark sets after a one-hour run. LLM-C consistently and significantly outperforms all baselines. On the synthetic suite, LLM-C achieves an average branch coverage of 91.3\%, compared to 75.6\% for classical concolic testing. The improvement is even more pronounced on the realistic fintech application, where LLM-C achieves 85.7\% coverage, demonstrating its effectiveness on real-world code.

Figure~\ref{fig:coverage} illustrates the time taken to reach 80\% branch coverage on the fintech application. LLM-C is substantially more efficient, reaching the target coverage in just 25 minutes. In contrast, the classical concolic tester requires 55 minutes, and the GA-based approach fails to reach 80\% coverage within the one-hour time budget. This highlights the ability of the LLM guidance to quickly focus the search on productive areas of the code.

\begin{table}[h]
  \caption{SMT Solver Invocations and Timeouts}
  \label{tab:solver}
  \begin{tabular}{lcc}
    \toprule
    Technique & Solver Invocations & Solver Timeouts \\
    \midrule
    Concolic & 15,432 & 1,234 \\
    \textbf{LLM-C} & \textbf{8,765} & \textbf{245} \\
  \bottomrule
\end{tabular}
\end{table}

Finally, Table~\ref{tab:solver} compares the number of SMT solver invocations and timeouts between the classical concolic tester and LLM-C. Our hybrid approach reduces the number of solver invocations by nearly 43\% and the number of timeouts by over 80\%. This is a direct result of the LLM-guided path prioritization, which avoids redundant and dead-end paths, and the constraint mutation mechanism, which helps to bypass difficult constraints.

\subsection{Analysis}

The experimental results strongly support our hypothesis that integrating LLMs into the concolic testing loop leads to a more efficient and effective testing process. The superior performance of LLM-C can be attributed to several factors. First, the LLM's semantic understanding of the code allows it to prioritize paths that are more likely to be interesting, thus avoiding the wasted effort of exploring trivial or redundant paths. Second, the ability of the LLM to assist with constraint solving significantly reduces the number of times the concolic engine gets stuck, allowing it to explore deeper into the program's state space. Finally, the semantic input synthesis provides a powerful fallback mechanism that can inject high-quality, domain-aware test cases into the process when symbolic exploration fails.

It is particularly noteworthy that the performance gap between LLM-C and the baselines is larger on the more realistic fintech application. This suggests that the semantic guidance provided by the LLM is even more valuable in the context of complex, real-world code with intricate business logic. While the overhead of the LLM API calls is not negligible, our results show that this cost is more than offset by the dramatic reduction in expensive solver invocations and the overall increase in testing efficiency.

\section{Threats to Validity}

We acknowledge several potential threats to the validity of our study.

\subsection{Internal Validity}

Threats to internal validity relate to potential errors in our experimental setup. The performance of our LLM-C prototype could be influenced by the specific implementation choices we made, such as the batching mechanism for LLM queries and the timeout values for the SMT solver. While we have attempted to tune these parameters to be fair, a different configuration might yield different results. Furthermore, bugs in our own prototype could have affected the outcomes.

\subsection{External Validity}

External validity concerns the generalizability of our findings. Our experiments were conducted on a limited set of benchmarks. While we included both synthetic programs and a realistic application, these may not be representative of all possible software. The performance of LLM-C could vary on different types of applications or programming languages. Additionally, our approach is tightly coupled with the specific LLM used (GPT-5.1). The performance may differ with other LLMs, and future changes to the model's API or behavior could impact our results.

\subsection{Construct Validity}

Construct validity relates to whether our metrics accurately measure what we intend to evaluate. We have used standard metrics like branch coverage and path coverage, which are widely accepted in the software testing community. However, high coverage does not necessarily equate to finding more bugs. While we believe that the increased exploration capabilities of LLM-C will lead to better bug detection, a more direct evaluation of its bug-finding effectiveness on a large corpus of real-world bugs is needed.

\subsection{LLM-Related Risks}

There are also risks inherent in the use of LLMs. The non-deterministic nature of some LLM outputs could make the testing process difficult to reproduce. The cost of using commercial LLM APIs could be a barrier to adoption for some users. Finally, there is a risk of the LLM developing biases based on its training data, which could lead it to ignore certain types of paths or bugs.

\section{Conclusion and Future Work}

In this paper, we have presented LLM-C, a novel hybrid software testing framework that integrates concolic execution with Large Language Models. Our approach leverages the semantic reasoning capabilities of LLMs to guide the path exploration process, resulting in a more efficient and effective testing methodology. The experimental results demonstrate that LLM-C significantly outperforms traditional concolic testing, as well as other baseline techniques, in terms of code coverage and testing efficiency. By combining the systematic exploration of concolic testing with the intelligent guidance of LLMs, our work opens up a new paradigm for automated software testing.

Our primary contribution is the demonstration that an LLM can serve as a powerful external heuristic to guide a concolic testing engine, overcoming long-standing challenges like path explosion and constraint solver bottlenecks. The ability of the LLM to reason about code at a semantic level allows the testing process to focus on interesting and bug-prone areas, moving beyond the limitations of purely syntactic heuristics.

The broader impact of this work lies in the potential for a new generation of AI-assisted software engineering tools. As LLMs become more powerful and accessible, their integration into development workflows will become increasingly common. Our work provides a concrete example of how this integration can be achieved in the critical domain of software testing, leading to more reliable and robust software.

For future work, we plan to explore several promising directions. First, we intend to conduct a larger-scale evaluation of LLM-C on a more diverse set of real-world applications and a comprehensive bug benchmark. Second, we will investigate the use of smaller, domain-specific language models, which could offer a more cost-effective and reproducible alternative to large, commercial LLMs. Finally, we plan to explore the application of our hybrid approach to other areas of software verification, such as security analysis and program repair.

\bibliographystyle{ACM-Reference-Format}
\bibliography{references}

@inproceedings{godefroid2005dart,
  author    = {Godefroid, Patrice and Klarlund, Nils and Sen, Koushik},
  title     = {{DART}: Directed Automated Random Testing},
  booktitle = {Proceedings of the 2005 ACM SIGPLAN Conference on Programming Language Design and Implementation (PLDI)},
  year      = {2005},
  pages     = {213--223},
  publisher = {ACM},
  address   = {New York, NY, USA},
}

@inproceedings{sen2005cute,
  author    = {Sen, Koushik and Marinov, Darko and Agha, Gul},
  title     = {{CUTE}: A Concolic Unit Testing Engine for {C}},
  booktitle = {Proceedings of the 10th European Software Engineering Conference held jointly with 13th ACM SIGSOFT International Symposium on Foundations of Software Engineering (ESEC/FSE)},
  year      = {2005},
  pages     = {263--272},
  publisher = {ACM},
  address   = {New York, NY, USA},
}

@misc{chen2022codet,
  title={Codet: Code generation with generated tests},
  author={Chen, Bei and Zhang, Fengji and Nguyen, Anh and Zan, Daoguang and Lin, Zeqi and Lou, Jian-Guang and Chen, Weizhu},
  journal={arXiv preprint arXiv:2207.10397},
  year={2022}
}

@inproceedings{tufano2019learning,
  author    = {Tufano, Michele and Watson, Cody and Bavota, Gabriele and Di Penta, Massimiliano and White, Martin and Poshyvanyk, Denys},
  title     = {Learning How to Mutate Source Code from Bug-Fixes},
  booktitle = {Proceedings of the 2019 IEEE International Conference on Software Maintenance and Evolution (ICSME)},
  year      = {2019},
  pages     = {301--312},
  publisher = {IEEE},
  address   = {USA},

}

@inproceedings{liu2023large,
  author    = {Liu, Yinlin and Xie, Chunqiu Steven and Deng, Yuxiang and Zhang, Lingming},
  title     = {Large Language Models are Zero-Shot Fuzzers: Fuzzing Deep-Learning Libraries via Large Language Models},
  booktitle = {Proceedings of the 32nd ACM SIGSOFT International Symposium on Software Testing and Analysis (ISSTA)},
  year      = {2023},
  pages     = {423--435},
  publisher = {ACM},
  address   = {New York, NY, USA},

}

@inproceedings{eslamimehr2014sherlock,
  title={Sherlock: scalable deadlock detection for concurrent programs},
  author={Eslamimehr, Mahdi and Palsberg, Jens},
  booktitle={Proceedings of the 22nd ACM SIGSOFT International Symposium on Foundations of Software Engineering},
  pages={353--365},
  year={2014},
  address={New York, NY, USA},
  publisher = {ACM},

}

@inproceedings{eslamimehr2015atomchase,
  title={AtomChase: Directed search towards atomicity violations},
  author={Eslamimehr, Mahdi and Lesani, Mohsen},
  booktitle={2015 IEEE 26th International Symposium on Software Reliability Engineering (ISSRE)},
  pages={12--23},
  year={2015},
  organization={IEEE},
  publisher={IEEE},
  address={Piscataway, NJ, USA}
}

@misc{JavaPathFinder_Repo,
  author       = {{NASA Ames Research Center}},
  title        = {{Java PathFinder (JPF) Core Repository}},
  howpublished = {\url{https://github.com/javapathfinder/jpf-core}},
  year         = {2026},
  note         = {GitHub repository. Accessed: 2026-01-17}
}

@misc{Z3_GitHub,
  author       = {{Microsoft Research}},
  title        = {{Z3 Theorem Prover}},
  howpublished = {\url{https://github.com/Z3Prover/z3}},
  year         = {2026},
  note         = {GitHub repository. Accessed: 2026-01-17}
}

@article{eslamimehr2014race,
  title={Race directed scheduling of concurrent programs},
  author={Eslamimehr, Mahdi and Palsberg, Jens},
  journal={ACM SIGPLAN Notices},
  volume={49},
  number={8},
  pages={301--314},
  year={2014},
  publisher={ACM New York, NY, USA}
}

\end{document}